# Orientation- and pressure-dependence of the vibrational response of a monolayer crystal on a vicinal diamond surface

Yi Zhao[1#], Lingxiao Zhao[1#], Chengjiang Du[1#], Ruirui Liu[1], John A. McGuire[1]*, Yanpeng Qi[1,2,3]*

1. School of Physical Science and Technology, ShanghaiTech University, 201210, Shanghai, China.
2. Shanghai Key Laboratory of High-resolution Electron Microscopy, ShanghaiTech University, 201210, Shanghai, China
3. ShanghaiTech Laboratory for Topological Physics, ShanghaiTech University, 201210, Shanghai, China

# These authors contributed to this work equally.

* Correspondence should be addressed to Y.Q. (qiyp@shanghaitech.edu.cn) and J.A.M.(jamcguire@shanghaitech.edu.cn).

**Abstract**: We systematically investigate polarization-dependent Raman spectra of a monolayer crystal of $WS_2$ on the (100) and (230) surfaces of diamond. At ambient pressure, identical polarization dependence of the Raman spectra is observed on the different surfaces, independent of the orientation of the monolayer crystal relative to the diamond crystal. However, when monolayer $WS_2$ is compressed to about 4 GPa, an abrupt drop of the intensity of the 2LA mode relative to the A' mode occurs on the (100) surface and the (230) surface with the zigzag direction along the atomic step edges of the (230) surface. In contrast, no such drop is observed when the armchair direction is along or at 15° to the atomic steps of the (230) surface. We also observe a shift of the polarization angle of the intensity maxima of the 2LA and A' modes on (230) surface during compression. These results demonstrate that the atomic steps of a vicinal surface strongly modify the vibrational response under high pressure.

**Keywords**: 2D Materials, High Pressure, Second Harmonic Generations, Raman Spectroscopy, Vibrational Properties, Vicinal Surfaces, Stacking Orientations

**Introduction**

Since the successful exfoliation of single-layer graphene, two-dimensional (2D) materials have attracted attention for their unique physical properties[1-5]. High electronic and thermal conductivity and excellent mechanical properties make 2D materials ideal for many applications[6-10]. However, most practical applications and fundamental studies of these low-dimensional materials entail placing or growing the samples on substrates at which point layer-substrate interactions can become important.

Layer-substrate interactions have been studied extensively in the literature. Xiaoli Hu et al. ascribed substrate-dependent graphene/tip contact conductances to the different elasticities of the substrates[11]. When 2D materials are grown by molecular beam epitaxy (MBE) or chemical vapor deposition (CVD), substrates are chosen to match the crystalline unit cell of the target material[12-14]. Band hybridization at the interface between monolayer $WSe_2$ (point group $D_{3h}$) and black phosphorus (point group $D_{2h}$) leads to symmetry breaking and the consequent emergence of a shift current.[15] Similarly, band hybridization at the interface between even-layer $MnBi_2Te_4$ (point group $D_{3h}$) and black phosphorus leads to symmetry breaking and the emergence of a quantum metric nonlinear Hall effect.[16]

Compared to conventional methods of modulating the properties of layered materials, like gating[17, 18], doping[19, 20], intercalation[17, 21-23] and magnetic or superconducting proximity effects[24-26], high pressure not only modifies the electronic, phonon or crystalline structures[27-31] of the layers themselves but also amplifies the layer-substrate interactions. Han et al. found that splitting of Raman peaks and sudden softening of all the Raman modes occurs at about 11 GPa for monolayer $WS_2$ on $Si/SiO_2$ substrate, but these phenomena were absent for monolayer $WS_2$ placed directly on a diamond surface[32]. However, reported high pressure *in-situ* Raman spectra exhibit inconsistencies[25, 31, 32]. For example, Negi et al. observed the disappearance of the 2LA modes at 8 GPa[33], which was preserved to at least 21.26 GPa in the work of Han et al. Moreover, to date, high-pressure studies of 2D materials have either focused on low-index surfaces or failed to note whether a low-index or vicinal surface was used.

The atomic steps of vicinal surfaces have the potential to perturb a monolayer in ways distinct from a low-index surface. In this work, we systematically study the polarization-dependent Raman (PDR) spectra and the rotational anisotropy of second harmonic generation (RA-SHG) of monolayer $WS_2$ under high pressure on diamond culets with different crystalline faces. The results indicate that under relatively low pressure (< 10 GPa), even different surfaces of the same substrate materials can lead to significant differences in PDR and SHG patterns. In the case of vicinal surfaces, these effects are correlated with the orientation between the monolayer crystal and the atomic steps of the diamond surfaces. For samples on culets presenting the (100) surface or the (230) surface with zigzag orientation of $WS_2$ (zigzag edges parallel to the step edges of the vicinal surface), abrupt drops of the relative intensity of the 2LA and E' modes at about 4-5 GPa occur. In contrast, in the case of armchair orientation and 15° orientation on the (230) culet, no similar phenomenon is observed. Apart from changes in vibrational modes, the SHG patterns change during compression on the (230) culet and indicate that symmetry breaking occurs in all three orientations.

**Results**

We first measure the polarization-dependent Raman (PDR) spectra of monolayer $WS_2$ at ambient pressure. Several Raman modes are observed in the wave number range between 250 cm$^{-1}$ and 450 cm$^{-1}$ (Figure 1b). According to the $D_{3h}$ point group of monolayer $WS_2$ and previous DFT and experimental results[34-37], these Raman modes can be ascribed to the 2LA mode, E' mode, A' mode and other higher order scattering modes. The representation of the 2LA mode can be derived from the irreducible representations of the $D_{3h}$ point group: 2LA=LA·LA=E'·E'=A'+E' (xy plane).

The Raman tensors of the 2LA, E' and A' modes are

| | Raman Tensor | Atomic displacement patterns |
|---|---|---|
| 2LA | $\begin{pmatrix} a+b & b & 0 \\ b & a-b & 0 \\ 0 & 0 & c \end{pmatrix}$ | 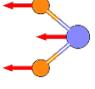 |
| E' | $\begin{pmatrix} 0 & d & 0 \\ d & 0 & 0 \\ 0 & 0 & 0 \end{pmatrix}, \begin{pmatrix} d & 0 & 0 \\ 0 & -d & 0 \\ 0 & 0 & 0 \end{pmatrix}$ | 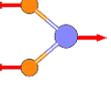 |
| A' | $\begin{pmatrix} e & 0 & 0 \\ 0 & e & 0 \\ 0 & 0 & f \end{pmatrix}$ | 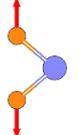 |

Raman spectra are collected in an epi-geometry. The incident field polarization is controlled by a half-wave plate with $\theta = 0°$ corresponding to a Z(X,Y)Z Raman configuration. According to the Raman tensors for the 2LA and A' modes, the maxima of the Raman intensity for the 2LA and A' modes should be at $\theta = 90°$, corresponding to a Z(X,X)Z Raman configuration, the ambient-pressure spectrum for which is shown in Figure 1b. The observed spectra can be fit well with Lorentzian functions. The intensities of the 2LA and A' modes from fitting of the PDR spectra are shown in Figure 1c. Both of the Raman modes' intensities can be fit well with a $\sin^2\theta$ function, consistent with the above theoretical expectation.

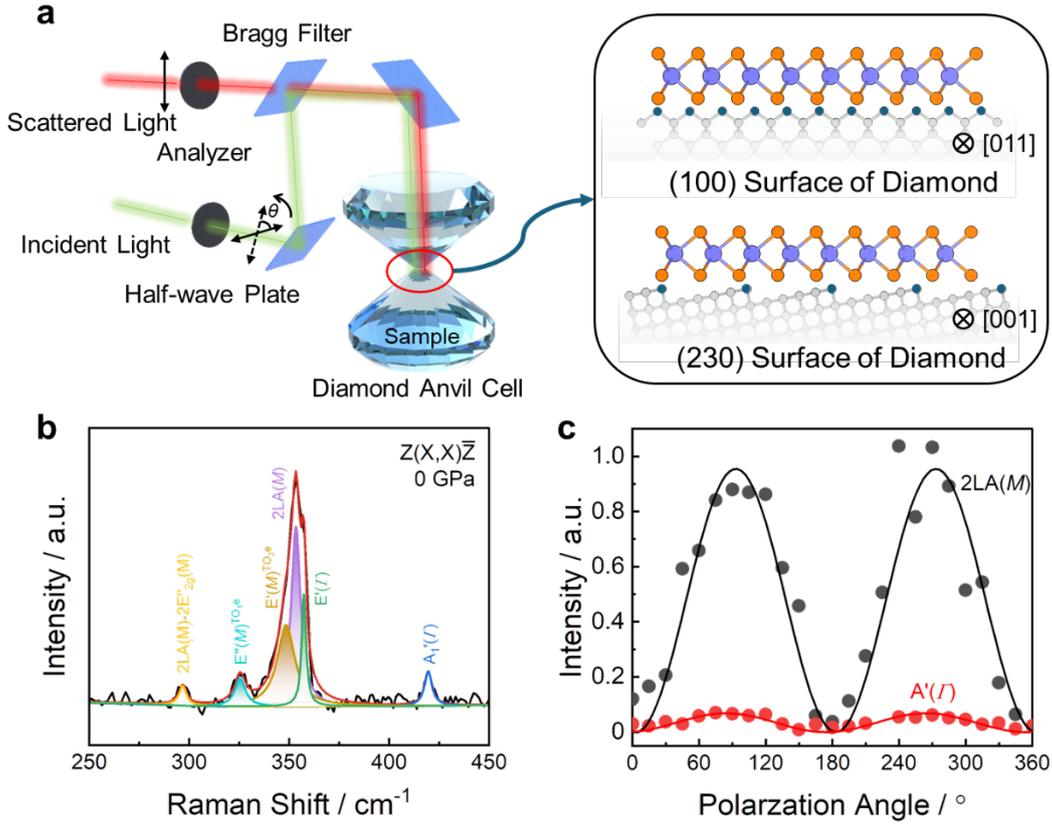

Figure 1. The schematic diagram of the high-pressure PD-Raman measurements with DAC. (a) The setup of optical devices. The interfaces between monolayer sample and the upper diamond culets of the DAC are shown in the inset. The violet and orange balls stand for W and S atoms respectively. (b) The PD-Raman spectrum at 0 GPa with the $Z(X, X)\bar{Z}$ configuration. The measured result is fitted by Lorenzian functions. (c) The fitted polarization-dependent intensity of the two peaks 2LA and A' modes. The solid lines show the fitting curve by a two-fold symmetric function.

To investigate the influence of the substrate on monolayer samples' vibrational properties under high pressure, we have measured the PDR spectra of monolayer $WS_2$ on different diamond culets with (100) and (230) crystalline faces. The (230) surface exhibits periodic atomic steps, so we also consider different $WS_2$ crystal orientations on the (230) surface: diamond step edges parallel to the $WS_2$ armchair edge (armchair orientation), the zigzag edge (zigzag orientation) or 15° rotated from the armchair or zigzag direction (15° orientation) of the monolayer $WS_2$.

High-pressure *in-situ* Raman spectra of monolayer $WS_2$ on a diamond (100) culet are presented in Figures 2a-c and S2. At ambient pressure (Figure 3a), the polarization dependences of the Raman peaks are identical to previously reported results[38, 39]. All the observed Raman modes exhibit two-fold rotational symmetry. The experimental intensity maxima of the A' and 2LA modes lie at $\theta = 90°$, as expected based on the Raman tensor and the geometric configuration of the experiment. However, the E' mode shows a maximum at about 0°. As shown in Figures 2b and c, the frequencies of the observed Raman modes increase gradually under increasing compression. The relative intensities of the Raman modes also change significantly. The intensity of the A' mode is enhanced compared with the 2LA mode at 5.0 GPa, while the two peaks at about 295 and 325 cm$^{-1}$ are no longer resolved. The slight shift of the angular location of the PDR maximum at different pressures is an artefact of errors in aligning the starting polarization direction. To confirm the symmetry of the electronic response of monolayer $WS_2$ on the (100) culet, we have also conducted SHG measurements. The results are shown in the insets of Figure 3a-c, which clearly show that the six-fold symmetry of the SHG patterns remain upon compression of the $WS_2$ on the (100) surface.

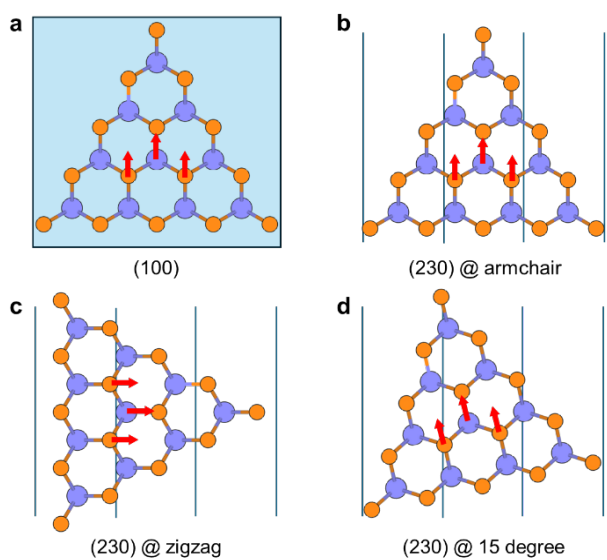

Figure 2. Schematic diagram of the interface between monolayer $WS_2$ and diamond culet as viewed from the z direction. (a) (100) culet, (b) (230) culet with armchair orientation, (c) (230) culet with zigzag orientation and (d) (230) culet with 15° orientation. The red arrows denote the direction of the atoms' displacements for the 2LA vibrational mode.

The Raman spectra of monolayer $WS_2$ on the (230) surface of diamond shown in Figures 2d-f and 3a-f are significantly different than those of $WS_2$ on the (100) surface. We first consider the case of armchair orientation (Figure 2d-f). Although the PDR spectra are identical to those for the (100) surface at ambient pressure, the relative intensities of the Raman modes are completely different, especially at high pressures (Figure 3f). Apart from that, the frequencies of the 2LA and A' modes are lower than those on the (100) face. We also note that the maximum polarization angle of the 2LA mode and A' mode significantly shifts during compression, indicating the deviation from the ideal Raman tensor for a system belonging to the $D_{3h}$ point group, which is discussed in detail in Supporting Information (Figure S6). The symmetry of the SHG pattern also changes at corresponding pressures (insets of Figure 3d-f).

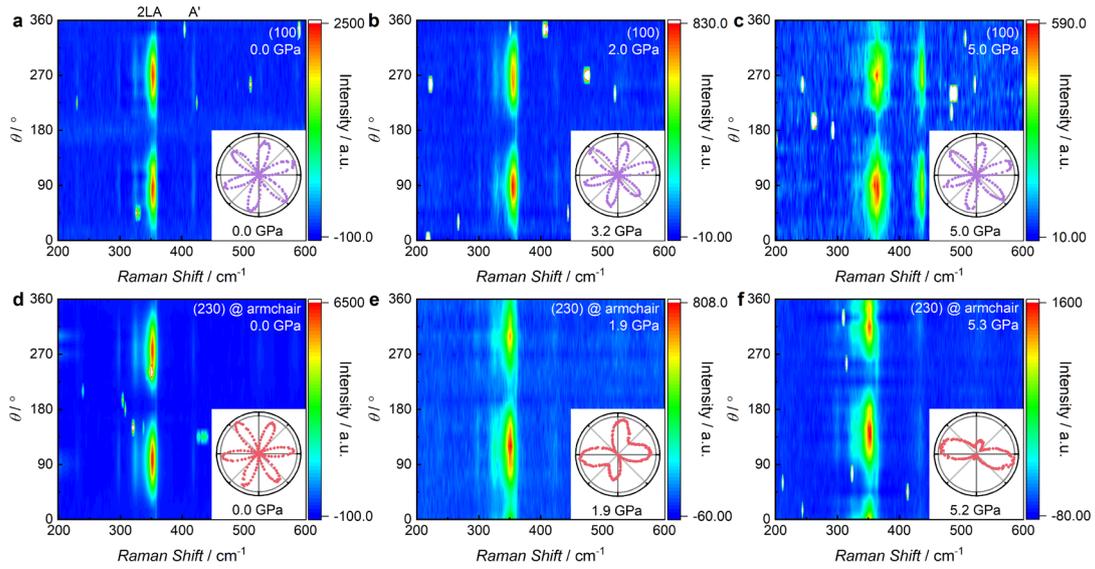

Figure 3. The PDR spectra of monolayer WS$_2$ under high pressure on a (100) culet and a (230) culet with armchair orientation. (a)-(c) (100) PDR spectra on (100) culet at ambient pressure, 2.0 GPa and 5.0 GPa. The insets show the SHG patterns at ambient pressure, 3.2 GPa and 5.0 GPa, respectively. (d)-(f) PDR spectra of WS$_2$ on a (230) culet with armchair orientation at ambient pressure, 1.9 GPa and 5.3 GPa. The insets show the SHG patterns at the same or similar pressures.

To investigate whether the relative WS$_2$ crystal orientation influences the vibrational properties of monolayer WS$_2$ on diamond (230), we have also measured PDR spectra of samples with the zigzag and 15° orientations. As shown in Figures 3a and d, at ambient pressure, the relative crystal orientations have no influence on the PDR spectra, showing identical features as Figure 4a and d. With compression to 1.9 GPa, though the two orientations still show nearly identical relative intensities of the Raman modes, the maximum polarization angles are quite different with different orientations (fitted in Figure S6). When we compress the samples to 4.2 GPa (Figure 4c), more effects of the orientation emerge. Again, similar to the case on the (100) culet, a dramatic drop of the intensity of the 2LA mode relative to the A' mode is observed in zigzag orientation. In contrast, in the case of 15° orientation, the intensity of the 2LA mode relative to the A' mode is not significantly suppressed, compared to what is observed at lower pressures. Another similarity between the zigzag orientation on the (230) culet and the (100) culet is that the frequencies of the 2LA and A' modes also blueshift compared with the 15° orientation at a comparable pressure (5.3 GPa). In addition, dramatic changes of rotational anisotropy are also observed in the SHG for the two orientations, as shown in the insets of Figure 4a-f.

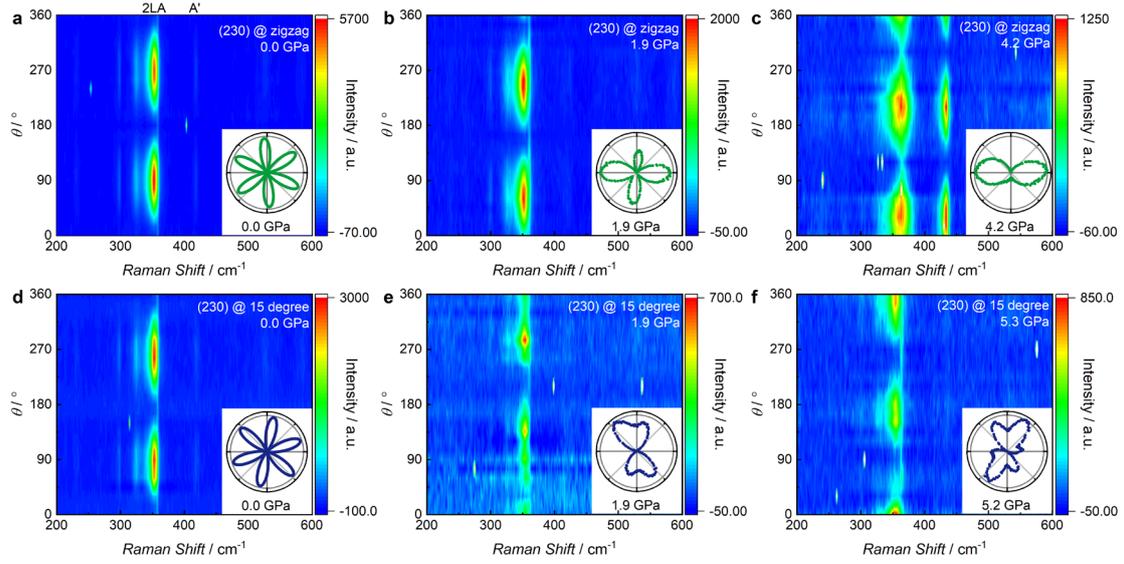

Figure 4. The PDR spectra of monolayer WS$_2$ under high pressure on a (230) culet with zigzag orientation and 15° orientation. (a)-(c) PDR spectra on a (230) culet with zigzag orientation at ambient pressure, 1.9 GPa and 4.2 GPa. The insets show the SHG patterns at the corresponding pressures. (d)-(f) PDR spectra on a (230) culet for 15° orientation at ambient pressure, 1.9 GPa and 5.3 GPa. The insets show the SHG patterns at the corresponding pressures, with the horizontal axis corresponding to the direction of the step edges.

**Discussions**

From the measured PDR spectra in Figures 2 and 3, we fit the results with Lorentzian functions. The fitted Raman shifts of the three main modes 2LA, E' and A' at different pressures are summarized in Figure 5a. The Raman shifts of the 2LA, E' and A' modes show different evolutions of the Raman shift under increasing pressure depending on the surface. Compared with the samples on the (230) culet, monolayer WS$_2$ on the (100) diamond culet shows larger Raman shifts. In tandem with the splitting of the Raman shifts, the relative intensity of the 2LA mode also decreases around 4 GPa on the (100) culet, as shown in Figure 5b. These two phenomena observed in the PD-Raman spectra indicate that the vibrations of the 2LA mode and the E' mode are suppressed on (100) culet or (230) culet with armchair orientation.

To understand the suppression of individual vibrational Raman features for certain surfaces and orientations, we consider the microscopic structure of WS$_2$ and the diamond (230) surface. Figure 2 shows schematic structures of the interface of diamond and monolayer WS$_2$. As shown in Figure 2a, the (100) face is atomically flat over extended distances and has no regular atomic-step-like structure. The 2LA and E' modes correspond to vibrations of W and S atoms along the armchair direction. The flat surface inhibits in-plane vibrations, especially under high pressure, since the interface interaction is enhanced, which applies a friction-like suppression to vibrations. In the case of the diamond (230) surface, these frictional forces are anisotropic. Under this scenario, the 2LA mode is significantly suppressed after compression to 4 GPa on the (100) diamond culet.

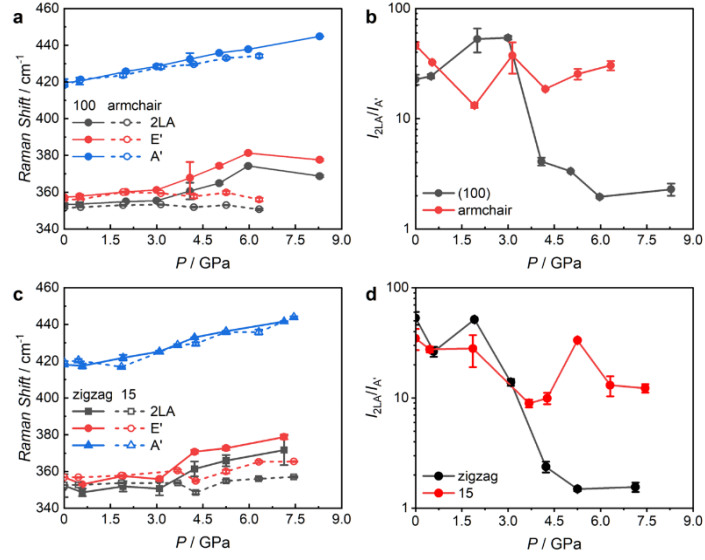

Figure 5. The vibrational properties of monolayer $WS_2$ on a (100) culet and a (230) culet with armchair orientation. (a) The pressure dependence of the fitted Raman shifts of the three Raman modes 2LA, E' and A' on (100) and (230) culets with armchair orientation. (b) The pressure dependence of the relative intensity $I_{2LA}/I_{A'}$ on (100) and (230) culets with armchair orientation. (c) The pressure dependence of the fitted Raman shifts on (230) culet with zigzag or 15° orientation. (d) The pressure dependence of the relative intensity $I_{2LA}/I_{A'}$ on (230) culet with zigzag or 15° orientation. The intensities at different pressure points are the integrals of the Raman peaks.

The orientations on the (230) culets are shown in Figures 2b-d. The diamond (230) surface can be characterized by periodic atomic steps spaced by about 6.43 Å with an overall periodicity of the surface structure of 12.86 Å with the top C atoms forming step edges along the *c* axis of the diamond crystal, which is parallel to the blue lines in Figure 2b-d. As a result of the incommensurate size between the $WS_2$ and diamond steps, most S atoms lie in the gaps between step edges and are expected to have weaker interactions with the surface carbon atoms during the vibrations associated with the 2LA mode in the armchair and 15° orientations. In contrast, in the case of zigzag orientation, the S atoms vibrate perpendicularly to the diamond steps. Vibration of the S atoms in this direction will face stronger obstruction by the substrate, thus exhibiting a suppression of the 2LA mode in PD-Raman spectra. These results indicate that the vicinal substrates can have pronounced orientation-dependent impact on the vibrations of 2D materials under high pressure.

The pressure-dependent RA-SHG of monolayer $WS_2$ on diamond (100) and (230) surfaces has been analyzed elsewhere by a bond-additivity model. RA-SHG at optical frequencies is a probe of the symmetry of the electronic structure. The fundamental photon energy used here is below the optical gap of $WS_2$, but the second harmonic is above the gap. The detailed second-harmonic response of $WS_2$ on the diamond (230) surface depends on where in reciprocal space the second-harmonic is resonant with the valence-to-conduction band transition and how the bands in these regions of reciprocal space are coupled to other regions of the band structure by a scattering potential with a wave vector corresponding to the period of the atomic steps of diamond (230). At the fundamental photon energy used in our SHG measurements, the dominant contribution to the SHG response is associated with the region around the $Q$ point (also called the $\Lambda$ point), which is near the midpoint between the $\Gamma$ and $K$ points of the Brillouin zone[40]. In contrast, Raman spectroscopy probes the Raman response of phonons or their overtones around the $\Gamma$ point. The 2LA Raman mode studied here is associated with the top of the LA phonon band around the $M$ point, where the dispersion is relatively flat. Likewise, the A' mode is a relatively localized mode excited around the $\Gamma$ point. Consequently, as the rotational symmetry of $WS_2$ on the diamond (230) surface is broken with the application of pressure, the evolution of the SHG and Raman rotational anisotropies are not expected to follow the same form, and the Raman response is expected to be sensitive to the local environment, which lends itself to an intuitive explanation for the pressure-induced suppression of this mode in the Raman response of $WS_2$ on diamond (230) under pressure.

**Conclusion**

In summary, we have systematically measured the PD-Raman spectra of monolayer $WS_2$ on different crystalline faces of diamond and with different relative crystal orientations under high pressure. The results show that for 2D materials, even though the Raman spectra at ambient pressure show negligible differences between different diamond surfaces and monolayer-diamond orientations, significant differences in Raman shifts and intensities may occur under high pressure, dependent on the substrate. For monolayer $WS_2$, the (100) culet or the (230) culet with zigzag orientation exhibits an abrupt drop of the intensity of the 2LA relative to the A' modes at about 4-5 GPa. In the armchair orientation and the 15° orientation on (230) surface, no similar phenomenon is observed. We conclude that high pressure has a strong effect in modulating the interactions between the 2D materials and the substrates, which are dependent on crystalline surface and orientation. These findings suggest careful choices of substrates for designing devices or investigations of 2D materials, especially under high pressure, and also unveil a new mechanism for modulating their vibrational modes.

**Experimental Methods**

Bulk $WS_2$ crystals are grown via chemical vapor transport[41]. Monolayer $WS_2$ samples are obtained by mechanical exfoliation of bulk samples. The Raman spectra in this work are measured by a Renishaw micro-confocal laser Raman spectrometer, with a 532nm, 50 mW laser light source. We use a half-wave plate to control the polarization of the incident laser, while the analyzer of the scattered light is held fixed perpendicularly to the starting direction of the half-wave plate. The beam splitter we use is a volume Bragg filter, which causes rather small anisotropy in the intensity of the incident beam. A 270 fs, 1030 nm laser pulses from an Yb:KGW laser operating at a repetition rate of 100kHz (CARBIDE, Light Conversion) were used to pump an OPA system (ORPHEUS-TWINS-F, Light Conversion) that generated the fundamental beam with a pulse duration of ~180 fs and an 800 nm center wavelength. The fundamental beam was passed through an achromatic half wave plate (350 nm-800 nm) and focused onto the samples in the DAC with a spot size of ~3 μm using a long-working distance ×20 objective under normal incidence. The reflected SHG signal was collected by the same objective, passed through the same half wave plate, reflected by a dichroic mirror, and passed through a linear analyzer oriented to pass light with polarization at the sample perpendicular to the fundamental polarization at the sample. A motorized rotation stage was used to rotate the half wave plate in 1° steps (2° steps of the linear polarization).

To apply high pressure, we utilize symmetrized diamond anvil cells with 400 um anvils. The monolayer WS2 samples are transferred onto the upper anvils, while the pre-indented and drilled iron gaskets are set on the lower anvils. Daphne oil 7373 is used as the pressure-transmitting medium and ruby fluorescence is used to calibrate the pressure in the sample chamber. For the upper diamond anvils which directly contact the monolayer WS2 samples, two diamonds with different crystalline faces (100) and (230) are utilized. During the transfer onto the (230) face, we have used three different orientations of the monolayer sample: armchair edge and zigzag edge parallel to or rotated 15° from the direction parallel to the step edges of the diamond.

**Acknowledgements**

This work was supported by the National Key R&D Program of China (Grant No. 2023YFA1607400), the National Natural Science Foundation of China (Grant No. 52272265) and start-up funds from Shanghaitech University (2017F0201-000-11). Samples preparation were supported by the Soft Matter Nanofab (SMN180827) and Analytical Instrumentation Center (Contract No. SPST AIC10112914), School of Physical Science and Technology, ShanghaiTech University.

**Supporting Information**

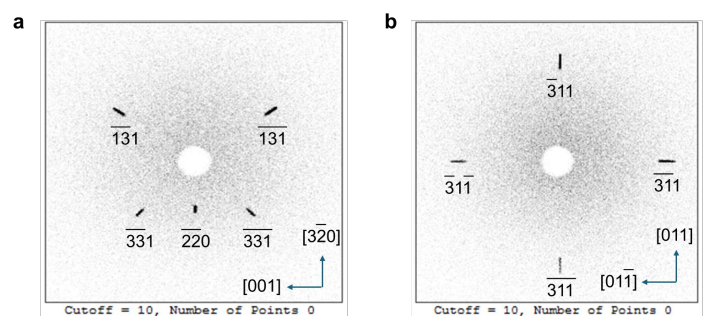

Figure S1. (a), (b) The Laue patten of diamond (230) and (100) surfaces.

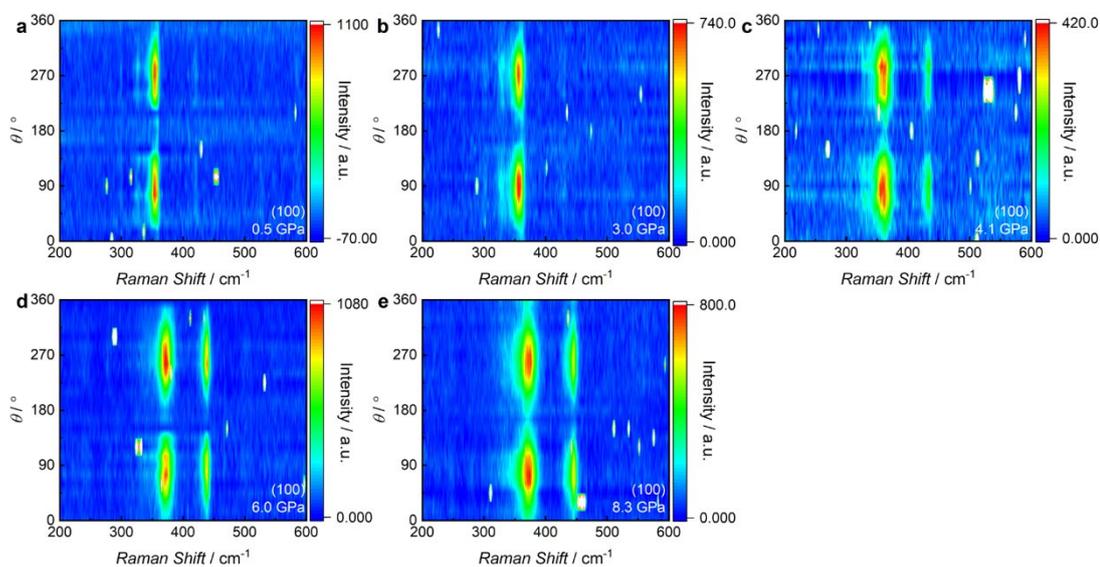

Figure S2. (a)-(e) The PD-Raman spectra of monolayer $WS_2$ under high pressure on the (100) culet.

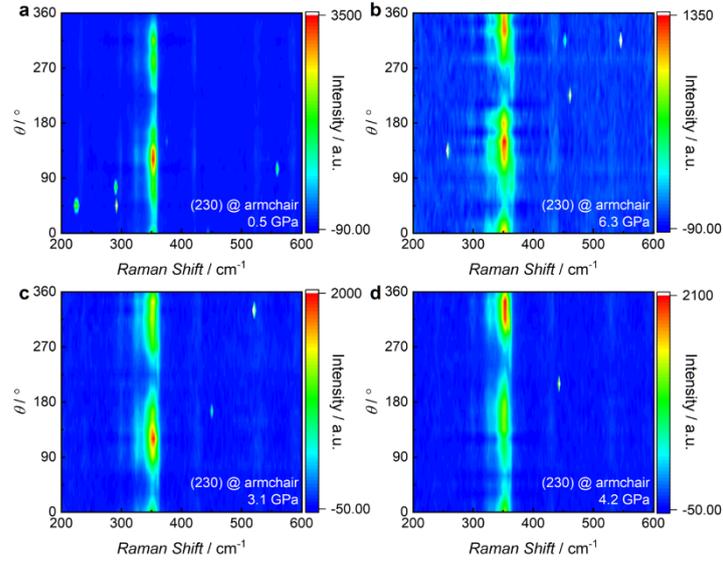

Figure S3. (a)-(d) PD-Raman spectra on the (230) culet with armchair orientation under different pressures.

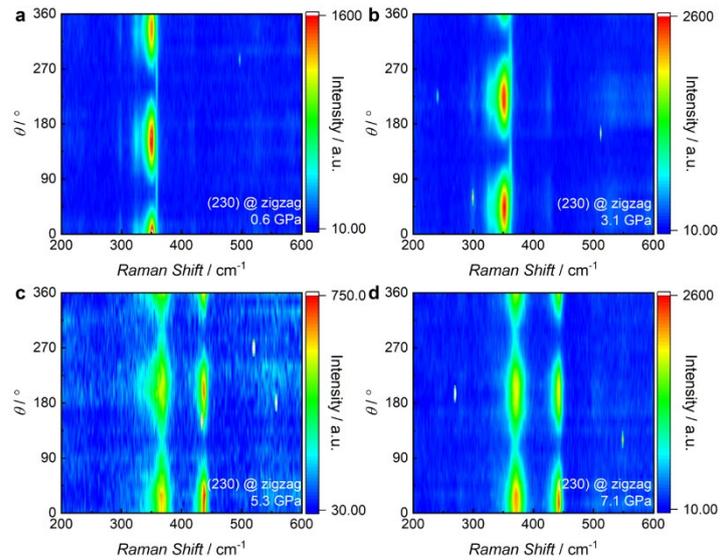

Figure S4. (a)-(d) PD-Raman spectra on the (230) culet with zigzag orientation under different pressures.

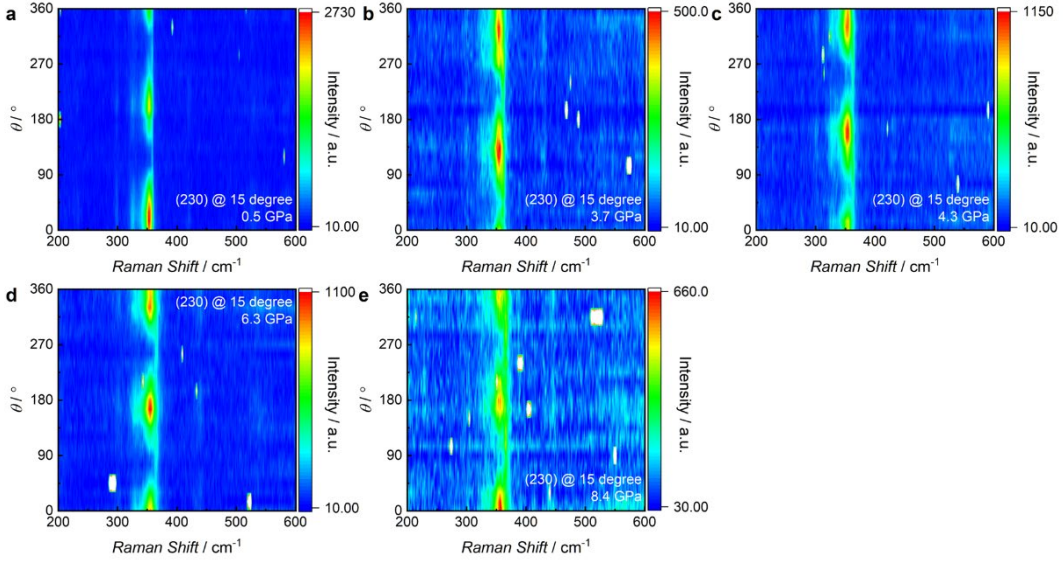

Figure S5. (a)-(e) PD-Raman spectra on (230) culet with 15° orientation under different pressures.

The scattered Raman intensity is given by $I = |e_i R e_s|^2$, where $e_i$ and $e_s$ are the polarization unit vectors of the incident and scattered radiation intensity, respectively, and $R$ is the Raman tensor. In the experimental setup, $e_i = (cos\theta, \ sin\theta)$ and $e_s = (0, 1)$. If we suppose that the actual Raman tensor has an arbitrary form $\begin{pmatrix} a & b \\ c & d \end{pmatrix}$, the intensity should be $I = (b \cos\theta + d \sin\theta)^2 = (A\sin(\theta + \phi))^2$, with $A^2 = b^2 + d^2$ and $tan(\phi) = \frac{b}{d}$. The polarization-dependent intensity of the A' mode under different pressures is fitted by a simple two-fold symmetric function $A^2 sin^2(\theta + \phi)$. The phase angle $\phi$ describes a rotation or shifting of the maxima of the PDR data during compression on the (230) culet. Further, the change in the tensor element ratio *b/d* is the physical origin of this rotation or shifting.

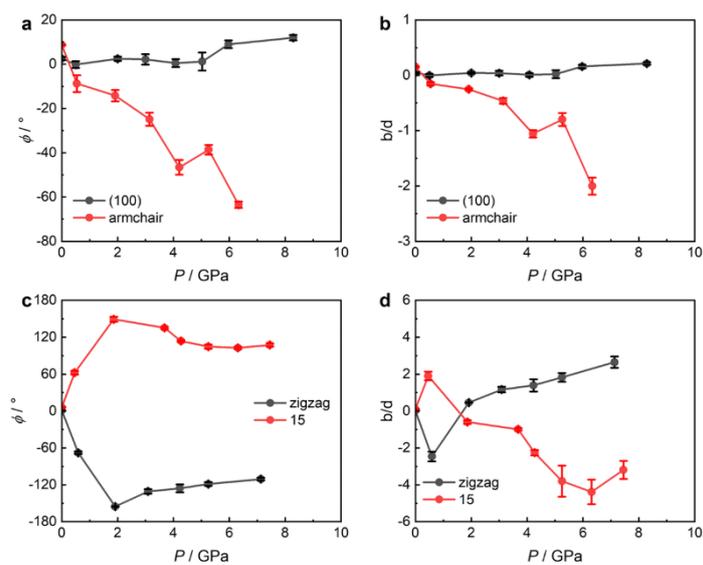

Figure S6. The fitted parameters of the A' mode on different diamond surfaces and orientations. (a) and (c) The fitted phase, ϕ, on the (100) surface and the (230) surface with armchair, zigzag or 15° orientations. (b) and (d) The corresponding fitted ratio of the tensor elements *b* and *d*.